\documentclass[floatfix,epsf,
 prb,twocolumn,
amsmath,amssymb,showpacs,superscriptaddress]{revtex4}
\usepackage[english]{babel}
\usepackage{graphics}
\usepackage{amsmath}
\usepackage{dsfont}
\usepackage{graphicx}
\usepackage[figtopcap]{subfigure}
\usepackage{enumerate}
\begin{document}\setlength{\unitlength}{1mm}
\newcommand{\ud}{\mathrm{d}}
\newcommand{\kvec}[2]{\begin{pmatrix} #1 \\ #2 \end{pmatrix} }
\newcommand{\rvec}[2]{\begin{pmatrix}#1 & #2 \end{pmatrix} }
\newcommand{\matt}[4]{\begin{pmatrix} #1 & #2 \\ #3 & #4 \end{pmatrix} }
\newcommand{\pdje}[1]{\partial_{#1}}
\newcommand{\ve}{\varepsilon}
\title{Edge magnetoplasmons in wide armchair graphene ribbons}

\author{O. G. Balev} 
\affiliation{Departmento de Fisica, Universidade Federal do Amazonas, 69077-000, Manaus, Brazil\\}
\author{P.~Vasilopoulos}
\affiliation{Department of Physics, Concordia University, 7141 Sherbrooke Ouest, Montr\'eal, Quebec, Canada H4B 1R6\\}

\author{H. O. Frota}
\affiliation{Departmento de Fisica, Universidade Federal do Amazonas, 69077-000, Manaus, Brazil\\}
\begin{abstract}
We show that near an armchair edge of a wide graphene channel, 
and in  the presence of a smooth step-like electrostatic lateral confining potential, 
the chirality, spectrum, spatial structure, and  number of the fundamental edge magnetoplasmons (EMPs), 
in  the $\nu=2$   regime of the quantum Hall effect, depend  strongly on the position of the Fermi level $E_{F}$. 
(i) When $E_{F}$ is small enough and intersects four degenerate states of the 
zero Landau level (LL) at one location and two degenerate states of this level at a different one,
two fundamental, counter propagating EMPs exist with opposite chirality.
This is in contrast with EMPs in conventional two-dimensional electron systems
in which only one fundamental EMP exists.
For the same wave vector these EMPs have  different moduli of phase velocities
and an essential spatial overlap.
These  EMPs can be on resonance in a wide range of frequencies, for micron or submicron lengths 
along the  edge.
(ii) When $E_{F}$ is sufficiently high and intersects only two degenerate states of the zero LL 
only one fundamental EMP exists with the usual chirality.
\end{abstract}
\maketitle
%

\section{Introduction}\label{sec1}
Since the experimental discovery of graphene and the manifestation of its  high-quality free-standing samples \cite{novo} ,
graphene has attracted a strong attention, see, e.g., Ref. \onlinecite{cast} for recent review.
It is currently the subject of many independent studies
because it's electronic properties are drastically different from
those, say, of conventional two-dimensional electron systems (2DES) in semiconductor
heterostructures, on liquid helium, etc.  Charge carriers in a single-layer graphene behave like "relativistic", chiral massless particles  with a
"light speed" equal to the Fermi velocity, $v_{F}$, and possess  a {\it
gapless,  linear} spectrum close to the $K$ and $K'$ points \cite{novo,cast,wallace}. 
One  major consequence is  the perfect transmission through arbitrarily high and wide barriers upon normal 
incidence, referred to as Klein tunneling \cite{cast,klein,kat}, and the direction-dependent  tunneling through barriers  \cite{kat}. 
Other unusual properties  are  a half-integer quantum Hall effect (QHE) \cite{cast,brey,aba,gus0}, 
a minimum metallic conductivity, and a zitterbewegung \cite{cast,zit1,zit2}. 
The latter effect is due to a lateral confinement of Dirac fermions 
and its manifestation can be essentially modified by a strong magnetic field \cite{zit1,zit2}.  
In addition, the submicron long mean-free paths \cite{novo} may have important consequences 
for applications in graphene-based devices, such as transistors, which have already been produced \cite{pom}.

In addition to the aforementioned studies, graphene's edges have also been studied considerably\cite{cast,brey,aba,gus0,loss,gus,milt},
in particular, in connection with the QHE \cite{cast,brey,aba,gus0}. For some properties  it matters whether the 
edges are of the armchair or zigzag type \cite{cast,brey,aba,gus0},  see  Refs.  \onlinecite{loss,gus} for nanoribbons. 
Also magnetic interface states (that can be understood as a new type of   edge states
made within a ''bulk'' of the graphene flake) in graphene-based quantum wires, created by a smooth
electrostatic confining potential, have been studied \cite{milt}.
There have also been many studies on plasmons \cite{plasmon,xue} and  magnetoplasmons \cite{magnplasmon},
collective wave excitations, in unconfined graphene; for a review on plasmons and magnetoplasmons in 
conventional 2DES, e.g., see \cite{volkov91,kushwaha2001}. 
Different types of edge magnetoplasmons (EMPs) have been studied 
theoretically \cite{fetter,volkov91,aleiner94,wen91,stone92,bal,bal2000,bal99,bal2000b} 
and experimentally  \cite{ashoori92,ernst96,kukushkin09} 
only for conventional 2DESs, see, e.g., Refs. \cite{volkov91,kushwaha2001}, and in particular Refs. \cite{ashoori92,ernst96,kukushkin09} 
for experimental studies of EMPs in the QHE regime. 
To our knowledge only the results of the EMP studies of Ref. \onlinecite{bal}, in the integral QHE 
regime, explain well  the experimental findings of \cite{ashoori92} and \cite{ernst96} for different 
realistic conditions. Indeed, the classical model of Ref. \onlinecite{volkov91}  
has major drawbacks in explaining the experimental results in the integer QHE  
regime, according to the analysis carried out in Ref. \onlinecite{ashoori92}.
This is also the case with the classical model of Ref. \onlinecite{aleiner94} 
according to Refs. \onlinecite{ernst96},  \onlinecite{kukushkin09}, and \onlinecite{bal}.

 In this work we explore theoretically the  possibility of fundamental EMPs in graphene  following  
Refs. \onlinecite{bal,bal2000}.   As will be shown, in  the presence of a smooth yet 
step-like lateral confining potential near 
 an armchair graphene edge, at $y=L_{y}/2$,  EMPs are possible in the $\nu=2$ QHE regime. These EMPs
 depend strongly on the position of the Fermi level $E_F$. 
For case (i), referred to in the abstract, the main resonance (Eq. (\ref{eq40}), of two EMPs of opposite chirality, 
localized within a submicron length of an armchair edge) is possible, e.g., if a strong coupling of the EMPs holds 
at the ends of the segment $L_{x}^{em} \leq L_{x}$, where $L_{x}$ is the length of the graphene channel. Our study shows that
the relevant condition $L_{x}^{em} \alt 1 \mu$m is realistic.
The EMPs that we find near a graphene edge are very different from those EMPs treated previously in  
conventional 2DES, in particular  in the integral QHE regime \cite{volkov91,aleiner94,wen91,stone92,bal,bal2000,bal99,bal2000b}.    

In Sec. II A we present the wave functions and the spectra of the Landau levels  (LLs) in an infinitely 
large graphene flake in the presence of a perpendicular magnetic field  and  of a smooth 
electrostatic confining potential, along the $y$ direction. 
In Sec. II B we study the combined effect of a  smooth, step-like electrostatic confining potential 
and of  armchair graphene edges, at $y=\pm L_{y}/2$,  on the local Hall conductivity 
in the $\nu=2$ QHE regime.  In Sec. III we present the resulting EMPs, at an edge region of a wide channel,  
and their strong dependence on  the position of  the Fermi level $E_F$. We make concluding remarks   in Sec. IV.

\section{Graphene Channel and local Hall conductivity}\label{sec2}
\subsection{Effect of a smooth potential on the LLs}

We consider an infinitely large flat graphene flake in the presence of a perpendicular magnetic 
field  ${\bf B}=B\hat{z}$ and  of a smooth confining potential $V_y=V(y)$ along the $y$ direction, of electrostatic
origin. For definiteness we assume that this potential is symmetric.
First we  consider solutions with energy and wave vector close to the K point.  In the nearest-neighbor, 
tight-binding model the  one-electron Dirac Hamiltonian, for massless electrons, 
is $\mathcal{H} = v_F \vec{\sigma}\cdot \hat{\vec{p}} + \mathds{1} V_y$, 
with $\mathds{1}$ the $2\times 2$ unit matrix. Explicitly $\mathcal{H} $ is given by ($e > 0$)
\begin{equation} 
	\mathcal{H} = v_F\matt{V_y/v_F}{p_x-i p_y-eBy}{p_x+i p_y-eBy}{V_y/v_F},
\label{eq1}
\end{equation}
where $p_x$ and  $p_y$ are the components of the  momentum operator ${\bf p}$  and $v_F \approx 10^6 m/s$ the Fermi velocity. 
The vector potential is taken in the Landau gauge, ${\bf A}=(-By,0,0)$.
The equation $(\mathcal{H} - E)\psi = 0$ admits solutions of the form 
\begin{equation}
	\psi({\bf r})= e^{i k_x x}\Phi(y)/\sqrt{L_x},\quad \Phi(y) = \kvec{A\Phi_A(y)}{B\Phi_B(y)} ,
\label{eq2}
\end{equation}
where $L_x$ is the length of the structure along the $x$ axis;
due to neglected spin-dependent contributions to the Hamiltonian Eq. (\ref{eq1})
each eigenstate is two times degenerated on the spin quantum number.  
The components $\Phi_A(y)$ and $\Phi_B(y)$ correspond to the two 
sublattices and are assumed normalized. Then the 
coefficients A and B satisfy the relation $|A|^2+|B|^2=1$. To simplify 
the notation in what follows we will write $\Phi_A(y)\equiv \Phi_A$ 
and $\Phi_B(y)\equiv \Phi_B$. Using Eqs. (\ref{eq1})  and (\ref{eq2}) we obtain
 \begin{eqnarray} 
&&\hspace*{-0.99 cm} 
V_y A\Phi_{A}+\hbar v_F\big( k_x-y/\ell_0^2 -\partial/\partial y\big)B\Phi_{B} =E 
A\Phi_{A},\\*
\nonumber
\ \\
&&\hspace*{-0.99 cm} \hbar v_F\big( k_x-y/\ell_0^2 +\partial/\partial y\big)A\Phi_{A} +V_y
B\Phi_{B}=EB\Phi_{B},
\label{eq3i4}
\end{eqnarray} 
where $\ell_0=(\hbar/eB)^{1/2}$ is the magnetic length. For $ E \neq V_y$ we 
solve Eq. (3) for $A\Phi_{A}$ and substitute the result in Eq. (4). Assuming  $B\neq 0$ and $E-V_y\neq 0$ this gives
%
\begin{eqnarray}  
\nonumber
\Big[\frac{\partial^2}{\partial \xi^2} &-&\xi^2 +\frac{\ell_0^2}{\hbar^2 v_F^2 }\big((E-U(\xi;y_{0}))^2
 +\frac{\hbar^2 v_F^2}{\ell_0^2}\big)\Big]\Phi_{B }(\xi)\\*
\nonumber
 \ \\
&-&\frac{d(E-U(\xi;y_{0}))/d\xi}{E-U(\xi;y_{0})}\,(\xi+\frac{\partial}{\partial \xi})\,\Phi_{B}(\xi)=0 ,
\label{eq6}	
\end{eqnarray} 
where we introduced the dimensionless variable $\xi=(y-y_0)/\ell_0$, with $y_0=\ell_{0}^2k_x$, 
and the notation $U(\xi;y_{0}) \equiv V(\ell_{0} \xi+y_{0})=V(y)$.

Assuming  $V_y$ is a smooth function of $y$, with a characteristic scale $\Delta y \gg \ell_{0}$,  we can make the approximation
\begin{equation} 
E-U(\xi;y_{0}) \approx  E-V(y_0)  -\xi \frac{\partial U(\xi;y_{0})}{\partial \xi}|_{\xi=0} ,
\label{eq7}	
\end{equation}
where it is used that $U(0;y_{0})=V(y_0)$.

First, for a given quantum number $y_{0}$, we assume $E=V(y_{0})$. 
Then we rewrite Eq. (\ref{eq6}), for $\xi\neq 0$ 
as  
\begin{equation}
\Big[\frac{\partial^2}{\partial \xi^2} -\frac{1}{\xi }\frac{d}{d\xi}-\xi^2 (1 -r^2)\Big]\Phi_{B }(\xi)=0,
\label{eq8}	
\end{equation}
where $a=\partial U(\xi;y_{0})/ \partial \xi|_{\xi=0}$ and  $r=\ell_0 a/\hbar v_F $.
Two independent solutions of Eq. (\ref{eq8}) are given by 
$\xi H^{(1)}_{1/2}(b\xi^2)$ and $\xi H^{(2)}_{1/2}(b\xi^2)$, with $b=(i/2)(1-r^2)^{1/2}$, 
where  $H^{(1,2)}_{1/2}(z)$
are the Hankel functions. 
The general solution is a linear combination of them.  
In particular, for a constant 
electric field of arbitrary strength, we can obtain exact results for the wave function and  eigenvalue of the $n=0$ LL
 for both $r < 1$ and 
 $r >1$, see Refs. \onlinecite{bask,per}. 

For the sake of comparison with the well-known exact results of Refs. \onlinecite{bask,per}, for a constant electric field
and $r < 1$,  we first assume  $r < 1$.
Then  we readily obtain $\Phi_{B }(\xi) \propto \xi H^{(1)}_{1/2}(\xi) \propto \exp[-(1-r^2)^{1/2}\xi^2/2]$, and  
%
\begin{equation} 
A\Phi_{A}(\xi) =(1/r) [1+(1/\xi)d/d \xi]B\Phi_{B}(\xi) ;
\label{eq9}	
\end{equation}
it follows that $\Phi_{A}(\xi) \equiv \Phi_{B}(\xi)$. So far the
calculations were performed for $\xi \neq 0$. 
Due to $E-U(0;y_{0})=0$, for $\xi=0$ 
the transition from Eqs. (3) and (4) to Eqs.(\ref{eq6}) and (\ref{eq8}) 
is not yet justified.  Here this special point
can be dealt with easily because the continuity of the wave function   
(\ref{eq2}) implies that  of $\Phi_{A}(\xi)$ and $\Phi_{B}(\xi)$ at $\xi=0$.

The procedure given so far applies to the K valley. If we repeat it for the K$^\prime$ valley, we obtain 
again Eq. (\ref{eq8}). If we label  the two valleys by $\kappa=\pm$, we can write both results
 in the form ($-\infty <\xi<\infty$) 
 \begin{eqnarray}  
&&\hspace*{-0.9cm}\Phi_{A\kappa }(\xi)=\Phi_{B\kappa }(\xi)=[(1-r^2)/\pi \ell_{0}^{2}]^{1/4}
e^{-(1-r^2)^{1/2}\xi^2/2}, \\*
&&A_\kappa=(1/r)[1-\kappa(1-r^2)^{1/2}]B_\kappa ,
\label{eq1011}	
\end{eqnarray}
where $\sqrt{2}A_\kappa=[1-\kappa(1-r^{2})^{1/2}]^{1/2}$ and 
$\sqrt{2}B_\kappa=r(1-\kappa(1-r^{2})^{1/2})^{-1/2}$. For $r\ll 1$ we have $B_+=A_-\approx 1$ and $A_+=B_-\approx r/2$.   
Notice that for the linear potential
$V_y=eEy$ the exact results of Refs. \onlinecite{bask} and \onlinecite{per} for the $n=0$ LL
coincide with ours, 
for the eigenvalue $E_{0,k_x}=eEy_0$,
and the wave function defined by Eqs. (9), (10). 
In what follows, we assume that $r \ll 1$.

Now, to study the effect of a smooth potential on the $n \neq 0$ LLs, we will use that for these LLs  $|E-V(y_0)| \gg |a|$. 
Then combining Eq. (\ref{eq7}) with Eq. (\ref{eq6}) we obtain  
\begin{equation}
\Big[\frac{\partial^2}{\partial \xi^2} -\xi^2 +\frac{ r^2}{a^2}\big[(E-V(y_0))^2 +\frac{\hbar^2 v_F^2}{\ell_0^2}\big]\Big]\Phi_{B k}(\xi)=0.
\label{eq12}
\end{equation}
This is a harmonic oscillator equation whose  solution is standard.  
For $N=1,2,...$ the eigenvalues are 
$E_{\pm N, k_x}=\pm (\hbar v_F /\ell_0)(2N)^{1/2} +V(y_0)$;   
the eigenfunctions are approximately the well-known ones for  
$V(y)=0$. We emphasize that as here $N \neq 0$, 
the condition $|E-V(y_0)| \gg |a|$ reduces to $r \ll 1$.

Finally, for  any $n=0, \pm 1,\pm 2,...$ LL  and $y_{0}$ not too close to the
graphene lattice termination at $y=\pm L_{y}/2$ (see Fig. 1 
which agrees with Ref. \onlinecite{milt}), the 
eigenvalues $E_{n, k_x}=E_{n, y_0}$ can be written as
\begin{equation}
E_{n, k_x}=sgn(n) (\hbar v_F/\ell_0)\sqrt{2|n|} +V(y_0),\quad n=0,\pm 1, \pm 2,...,
\label{eq13}
\end{equation}
where the sign function $sgn(n)=1$ and $-1$ for $n>0$ and $n<0$, respectively.
Notice that each $n \neq 0$ LL is twice degenerate with respect to the valley quantum number $\kappa$. 
Accordingly, for  any $n \neq 0$ LL  and $y_{0}$ not too close to the
graphene lattice termination at $y=\pm L_{y}/2$ (see Figs. 1(a), 1(b), compare   with Ref. \onlinecite{milt}) the 
eigenvalues (\ref{eq13}) are four times degenerate due to 
the  spin and valley degeneracies.

\subsection{Effect of a smooth potential and of an armchair edge on LLs\\
and local Hall conductivity in the $\protect\nu=2$ QHE regime}
\begin{figure}[ht]
\vspace*{-1.5cm}
\includegraphics [height=12cm, width=8cm]{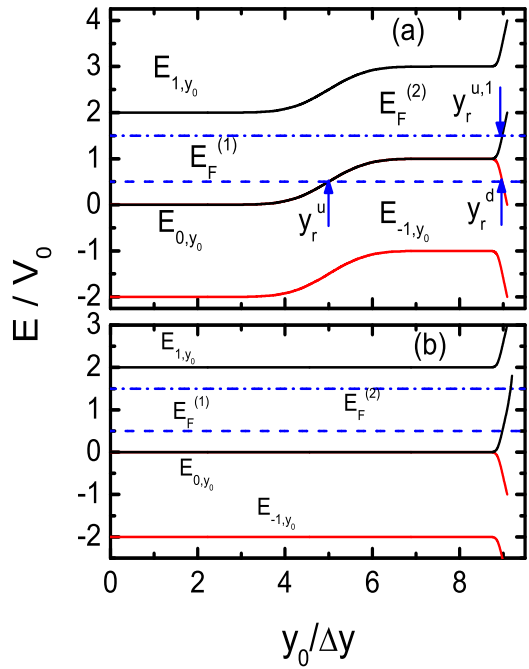}
\vspace*{-2.0cm}
\caption{(Color online) (a) Energy spectrum of the $n=-1$, $0$, and $1$ LLs 
as a function of the quantum number $y_{0}$, pertinent to the right half of a symmetric graphene 
channel with  armchair edges and the 
smooth electrostatic potential of finite strength, Eq. (\ref{eq17}), 
for two different  values of the Fermi level: (i) $E_{F}^{(1)}=V_{0}/2$ and 
(ii) $E_{F}^{(2)}=3V_{0}/2$.
For  both cases the $\nu=2$ quantum Hall effect is manifested in dc magnetotransport;
here 
$V_{0}=\hbar v_{F}/\sqrt{2} \ell_0$, 
$\Delta y=10\ell_{0}$, and $\ell_{0}$ is the magnetic length.
In case (i) two spatially separated edge states, of opposite chirality and different degeneracy, are
created at  $y_{r}^{u}=5\Delta y$ as the four times degenerate $n=0$ LL crosses 
the Fermi level $E_{F}^{(1)}$, and at $y_{r}^{d} \approx 9 \Delta y$ 
as the doubly degenerate branch of the $n=0$ LL crosses $E_{F}^{(1)}$.
Both $y_{r}^{u}$ and $y_{r}^{d}$ are marked by an upward arrow.
In case (ii) only the $n=0$ LL  at  $y_{r}^{u,1} \approx 9\Delta y$ (downward arrow) 
crosses the Fermi level $E_{F}^{(2)}$,
by its doubly degenerate branch that goes up with increasing $y_{0}$.
(b) The same LLs for a graphene channel in the 
absence of a smooth electrostatic potential; this spectrum  agrees with that of Refs. \onlinecite{brey,aba,gus0,gus}. 
Here for both  positions $E_{F}^{(1)}$ and $E_{F}^{(2)}$ of the Fermi
level  the  picture of the edge states is qualitatively the same as case (ii) of Fig. 1a.}
\end{figure}

Extending magnetotransport formulas for the local Hall conductivity $\sigma_{yx}(y)$ of a standard  2DES in the channel,
in the presence of a smooth, lateral confining potential \cite{bee,thouless93,bal96},   we obtain, for linear responses  
and in strong magnetic fields, $\sigma_{yx}(y)$ in the form \cite{zhe}
\begin{equation}
\sigma_{yx}(y)= n(y) e/B ,
\label{eq14}
\end{equation}
where $n(y)$ is the $y$-dependent electron density given by 
\begin{equation}
n(y)=  \sum_{\alpha\kappa}f_{\alpha\kappa}\langle\alpha\kappa| \mathds{1}\delta({\bf r}-\hat{{\bf r}})|\alpha\kappa\rangle , 
\label{eq15}
\end{equation} 
with $\alpha=\{n,k_x\}$. 
For  a finite hole density, $p(y)$, in Eq. (\ref{eq14}) it follows that $n(y)$ is changed
on $[n(y)-p(y)]$.
The standard relation $\sigma_{yx}(y)=-\sigma_{xy}(y)$ holds. 
Equation (\ref{eq14}) can be rewritten as
\begin{eqnarray}  
\nonumber
\sigma_{yx}(y)&=&2\frac{e^{2}}{ 
h}  \sum_{n \geq 0,\kappa=\pm} \int_{-\infty}^{\infty} dy_{0} [f_{n,y_{0},\kappa}-\delta_{n,0} \delta_{\kappa,-}] \\
&&\times [|A_{\kappa}^{n}|^2|\Phi_{A\kappa}^{n}(\xi)|^2+|B_{\kappa}^{n}|^2|\Phi_{B\kappa}^{n}(\xi)|^2],
\label{eq16}	
\end{eqnarray}
with $f_{n,y_{0},\kappa}$ the Fermi function and  
$A_{\kappa}^{0}$, $B_{\kappa}^{0}$, $\Phi_{A\kappa}^{0}(\xi)$,
$\Phi_{B\kappa}^{0}(\xi)$  given by Eqs. (10)-(11) in the linear-response limit $r \to 0$; the factor $2$ accounts for spin degeneracy. 
Point out that $\kappa=\pm$ in Eqs. (\ref{eq15}),(\ref{eq16}) is actually understood as the pseudospin 
quantum number; because, at $y_{0}>0$, only for 
$(L_y/2-y_0)/\ell_0 \gg 1$ it can be well approximated by the valley index.  
Indeed, a strong splitting between the electron, $\kappa=+$, and the hole, $\kappa=-$, branches
of the $n=0$ LL  \cite{cast,milt,brey}, due to 
hybridization of the valley states 
take place nearby the armchair edge, at $|L_y/2-y_0| \leq \ell_0$. The eigenvalues of the $n=0$ LL for 
$\kappa=+(-)$ increase (decrease) with increasing $y_{0}$;
in Eq. (\ref{eq16}) it is used that the contribution from the $(n=0,\kappa=-)$ LL is described better by the hole representation.
However, for the $n \geq 1$ LLs the $\kappa=\pm$ branches at the armchair edge
have a small splitting, due to hybridization of the valley states, as their eigenvalues increase with increasing $y_{0}$;
these branches are attributed to the electron band.

We now consider the situations depicted 
in Fig. 1(a) for a wide symmetric channel $L_{y} > 2y_r \gg \Delta y\gg \ell_0$, where $L_y/2=9 \Delta y$,
$y_r/\Delta y=5$, and $\Delta y/\ell_{0}=10$. However, $y_r/\Delta y$, and $\Delta y/\ell_{0}$
can take any large value if the EMPs at the right part of the channel 
are well decoupled from those at its left part. For clarity the smooth lateral potential is taken as 
\begin{equation}
\hspace*{-0.2cm}V(y)= 
(V_0/2)\Big[2+\Phi ((y-y_r)/\Delta y) +\Phi ((y+y_r)/\Delta y)\Big],
\label{eq17}
\end{equation}
where $\Phi (x)$ 
is the probability integral. 
When the Fermi level $E_F$ is between the bottoms of the $n=0$ and $n=1$ LLs, 
at $y_{0}=0$, and the condition $V_0\gg 2k_BT$ holds, the occupation of the $n \geq 1$ LLs is negligible; the same holds for  
the $n=0$ LL in the regions of $y_{0}$ that are well above $E_F$, 
 see Fig. 1(a). In addition to the smoothness of the potential  
 (\ref{eq17}), we assume armchair edges of the graphene sheet at $y=\pm L_y/2$, 
which cause the bending of the  LLs, \cite{cast,brey,aba,gus0,gus,milt}. 
In Fig. 1(a) we have $L_y/2-y_r=4 \Delta y$ but our main  results  hold 
qualitatively for $L_y/2-y_r \agt \Delta y$ as well. 

We point out that the fine structure of the LLs $n=\pm 1$ in Figs. 1(a), (b) , due to the removal 
of the pseudospin degeneracy at $|L_y/2-y_0| \leq \ell_0$ resulting from 
two possible hybridizations of the valley states for $n \neq 0$ LLs \cite{brey,aba}, 
is discarded. However,  a strong splitting of the $n=0$ LL 
 \cite{cast,milt,brey} at $|L_y/2-y_0| \leq \ell_0$, due to 
only one possible hybridization of the valley states 
\cite{brey,aba}, is taken into account. 
Due to these assumptions,  for $y_{0}$ very close to the armchair 
termination we can formally assume that the eigenvalues in Eq. (\ref{eq16}) and the wave functions
of the $n \neq 0$ LLs are independent of $\kappa$ while
the eigenvalues of the $n=0$ LL, for $\kappa=+(-)$ 
increase (decrease) with increasing $y_{0}$.

 For either  Fermi level position, $E_{F}^{(1)}$ or $E_{F}^{(2)}$ in Fig. 1(a), the $\nu=2$ quantum Hall 
regime will be manifested in  dc transport measurements. 
Indeed, for the $x$ axis normal  to the plane of Fig. 1(a) and magnetic fields $B>0$, it follows that at  $y_{0}=y_{r}^{u}$ ( 
$y_{r}^{u}=y_{r}=5\Delta y$)  the fourfold degenerate $n=0$ LL 
(composed both of a doubly degenerated conduction band branch, $\kappa=+$, and a doubly degenerated valence band branch, $\kappa=-$) crosses 
the Fermi level $E_{F}^{(1)}$ as it goes up, with increasing $y_{0}$. 
This creates a fourfold degenerate edge state that propagate 
along the {\it positive}  $x$ axis. However, at $y_{0}=y_{r}^{d}$ (here $y_{r}^{d} \approx 9 \Delta y$ is very close to the 
armchair termination of the graphene channel) only a doubly degenerated valence band branch ($\kappa=-$, with different spin quantum numbers
but the same hybridization of the valley states) of the
$n=0$ LL goes down  and crosses $E_{F}^{(1)}$ with increasing $y_{0}$. 
This branch creates a doubly degenerate edge state that propagate along the  {\it negative}  
$x$ axis. We call this situation case (i).
 For the Fermi level position $E_{F}^{(2)}$  in Fig. 1(a) only a doubly degenerated conduction band branch
($\kappa=+$, with different spin quantum numbers and the same hybridization of the valley states) of the
$n=0$ LL goes up with increasing $y_{0}$  and crosses $E_{F}^{(2)}$ at  $y_{0}=y_{r}^{u,1}$, where 
$y_{r}^{u,1} \approx 9 \Delta y$ is very close to the armchair termination of the graphene channel. 
This branch creates a doubly degenerate edge state that propagate along the  {\it positive}  
$x$ axis. We call this situation case (ii).

In Fig. 1(b) we plot the same LLs of the graphene channel in the 
absence of a smooth electrostatic potential. The spectrum shown is 
in agreement with that of Refs. \onlinecite{brey,aba,gus0,gus}. 
Here for both $E_{F}^{(1)}$ and $E_{F}^{(2)}$ the picture of the edge states 
is qualitatively the same as that for  case (ii) of Fig. 1(a) and 
magnetotransport measurements will manifest the $\nu=2$ QHE. 
Below we show that in cases  (i) and (ii)
the properties of EMPs are very different.

In   case (i), for $y_{0}>0$ and $(y_{r}^{d}-y_{0})/\ell_{0} \gg 1$, from 
Eqs. (9)-(10), (\ref{eq13}), (\ref{eq16}) and (\ref{eq17}) we obtain  
\begin{equation}
\sigma_{yx}(y)=\frac{4e^2}{h}\Big[1+\exp\left([V(y)-V(y_r^u)]/k_BT\right)\Big]^{-1}-\frac{2e^2}{h} ,
\label{eq18}
\end{equation}
where it is assumed that $E_{0,y_{0}}=V(y_{0})$ is smooth
on the  scale of $\ell_{0}$, i.e.,  $\ell_{0} dV(y_{r}^{u})/dy \ll k_{B} T$; 
the factor $4$ accounts  for spin and pseudospin degeneracy. 
This condition of smoothness can be rewritten, upon introducing 
the characteristic length $\ell_{T}=\ell_0(k_BT\ell_0/\hbar v_g^{u})$, 
as $\ell_{0} \ll \ell_{T}$, where $v_{g}^{u}=\ell_{0}^{2}\, \hbar^{-1} dV(y_{r}^{u})/dy$ 
is the group velocity at the edge $y_{r}^{u}$. Notice that  
by using $V_{0}=\hbar v_{F}/\sqrt{2} \ell_{0}$, $\ell_{0}/\Delta y=0.1$, and
all other conditions applying to Fig. 1, we obtain 
$v_{g}^{u}=(\ell_{0}/\sqrt{2 \pi} \Delta y) \times v_{F} \approx 4 \times 10^{6}$\,cm/s. 
Also, for qualitatively similar conditions we obtain   
$v_{g}^{u}/v_{F}=(\ell_{0}/\sqrt{2 \pi} \Delta y) \lll 1$ due to the condition
$\ell_{0}/\Delta y \ll 1$.

For sufficiently smooth potentials we can write
\begin{equation}
V(y_r^u+(y-y_r^u))\approx V(y_r^u)+(y-y_r^u) dV(y)/ dy|_{y=y_r^u} ,
\label{eq19}
\end{equation}
where the second term can be written as 
\begin{equation} 
(\hbar/\ell_{0}^2)\,(y-y_r^u)\frac{dE}{\hbar\, dk_x}|_{y=y_r^u} = 
(\hbar/\ell_{0}^2)\,v_g^{u} (y-y_r^u) .
\label{eq20}
\end{equation}
For  $|y-y_r|\leq \ell_T$ the approximation (\ref{eq19}) and Eq. (\ref{eq18}), 
for $(\ell_T/\Delta y)^2\ll 1$,  allow us to rewrite Eq. (\ref{eq18}) as
\begin{equation}
\sigma_{yx}(y)=\frac{4e^2}{h}
\Big[1+\exp[(y-y_r^u)/\ell_T]\Big]^{-1}-\frac{2e^2}{h}.
\label{eq21}
\end{equation}
 We remark that setting $\bar{y}=y-y_{r}^{u}$ gives  \cite{bal2000}
\begin{equation}
\frac{d\sigma_{yx}(y)}{dy}=
-\frac{4e^2}{h} \frac{1}{4\ell_T cosh^{2}(\bar{y}/2\ell_{T})}.
\label{eq22}
\end{equation}
Hence,  in case (i),  for $y>0$ and $(y_{r}^{d}-y)/\ell_{0} \gg 1$, we have 
Eqs. (\ref{eq21})-(\ref{eq22}). Further, for $L_{y}/2 \geq y \geq L_{y}/2-5\ell_{0}$ 
we model the $\nu=2$ numerical results \cite{brey,milt,gus,aba} with  the density
\begin{equation}
\hspace*{-0.2cm}n(y)-p(y)=\frac{1}{\pi^{3/2}\ell_0^3}\int_{-\infty}^{\infty}dy_0 
\,e^{-(y-y_0)^2/ \ell_0^2} \left[f_{0,y_0,-}-1\right] ,
\label{eq23}
\end{equation}
where we assumed that $E_{0,y_0,-}$  
is a sharply decreasing function at $y_{0} \approx y_{r}^{d}$ such that
the Fermi function in Eq. (\ref{eq23}) is very fastly growing at $y_{0} \approx y_{r}^{d}$ on a scale smaller
than $\ell_{0}$. Then from Eqs. (\ref{eq14}) and  (\ref{eq23}) we obtain
\begin{equation}
d\sigma_{yx}(y)/dy=(2e^2 /h\sqrt{\pi}\ell_0)\, e^{-(y-y_r^d)^2/\ell_0^2} ,
\label{eq24}
\end{equation}
by changing the derivatives with respect $y$ to those  
with respect $y_{0}$ and integrating by parts.

In a similar manner, for  case (ii) and  $y>0$, we obtain that $E_{0,y_0,+}$ is 
a sharply increasing function at $y_{0} \approx y_{r}^{u,1}$ and
\begin{equation}
d\sigma_{yx}(y)/dy  =-(2e^2  
/h\sqrt{\pi}\ell_0)\, e^{-(y-y_r^{u,1})^2/\ell_0^2} ,
\label{eq25}
\end{equation}
in agreement with Fig. 1(a).

\section{Strong dependence of EMPs on the Fermi-level position for $\protect\nu=2$}

Now we will study EMPs for  cases (i) and (ii), see Fig. 1, neglecting   dissipation. 
We expect that the charge excitation due to EMPs at 
the right part of channel will be strongly localized
 at $y_{r}^{u}$ ($\rho^{ru}(t,{\bf r})$) and $y_{r}^{d}$ ($\rho^{rd}(t,{\bf r})$), in  case (i), 
and at $y_{r}^{u,1}$ ($\rho^{r,u1}(t,{\bf r})$) in case  (ii). Then for  case (i)  
 the components of the  current density ${\bf j}(y)$, in the 
low-frequency limit $\omega\ll v_F/\ell_0$, are \cite{bal2000,bal} 
\begin{equation}
\hspace*{-0.25cm}j_x(y)=-\sigma_{yx}E_y(y)+v_g^u\rho^{ru}(\omega, k_x,y)+v_g^d\rho^{rd}(\omega, k_x,y),
\label{eq26}
\end{equation}
\begin{equation}
j_y(y)=\sigma_{yx}(y)E_x(y),
\label{eq27}
\end{equation}
where we suppressed the factor $\exp[-i(\omega t-k_xx)]$ common to all terms in Eqs. (\ref{eq26}) and (\ref{eq27}).
From  Eqs. (\ref{eq26}) and (\ref{eq27}), Poisson's equation, and the linearized continuity equation we find the integral equation for
the charge density $\rho(\omega, k_x,y)=\rho^{ru}(\omega, k_x,y)+\rho^{rd}(\omega, k_x,y)$  
\begin{eqnarray}
\nonumber
&&\hspace*{-0.99cm}(\omega -k_xv_g^u)\rho^{ru}(\omega, k_x,y)+(\omega -k_xv_g^d)\rho^{rd}(\omega, k_x,y)\\*
\nonumber
\ \\
\nonumber
&&\hspace*{-0.99cm}+{2k_x\over\epsilon} {d\sigma_{yx}(y)\over dy}\int_{-\infty}^{\infty} dy'\,
R_{g}(|y-y^{\prime}|,k_{x};d) 
 \\*
&&\hspace*{-0.99cm}
\times \left[\rho^{ru}(\omega, k_x,y')+\rho^{rd}(\omega, k_x,y')\right]=0.
\label{eq28}
\end{eqnarray}
For a metallic gate  placed on top of the sample, at a distance $d$ from the 2DES 
(usually  this is a heavily doped Si separated from the graphene sheet by a SiO$_{2}$ 
layer of  thickness $d=300$ nm),  $R_g(...)$ is given by
\begin{eqnarray}
\hspace*{-0.89cm} R_{g}(|y-y^{\prime}|,k_{x};d)&=&K_{0}(|k_{x}||y-y^{\prime }|) \nonumber \\
&-& K_{0}(|k_{x}|\sqrt{(y-y^{\prime })^{2}+4d^{2}}),    
\label{eq29b}
\end{eqnarray}
where $K_0(x)$ is the modified Bessel function. In the absence of a metallic 
gate, $d \to \infty$, the   dielectric constant $\epsilon $ is
spatially homogeneous if not stated otherwise.

As $d\sigma_{yx}(y)/dy$ is  too small according to Eqs. (\ref{eq22}) and (\ref{eq24}) 
except at $y \approx y_{r}^{u}$ and $y_{r}^{d}$, we rewrite Eq. (\ref{eq28}) as
\begin{eqnarray}
\nonumber
&&(\omega -k_xv_g^u)\rho^{ru}(\omega, k_x,y)+(\omega -k_xv_g^d)\rho^{rd}(\omega, k_x,y)\\*
\nonumber
&&- 
c_{h}k_x\left[\frac{1}{2\ell_T cosh^{2}(\bar{y}/2\ell_{T})}
-\frac{1}{\sqrt{\pi}\ell_0} e^{- (y-y_r^d)^2/\ell_0^2}\right]
\\*
\nonumber
&&\times \int_{-\infty}^{\infty} dy' 
R_{g}(|y-y^{\prime}|,k_{x};d) 
\\*
&& \times \left[\rho^{ru}(\omega, k_x,y')+\rho^{rd}(\omega, k_x,y')\right]  =0,
\label{eq29}
\end{eqnarray}
where $c_{h}=4e^{2}/h \epsilon$. In the long-wavelength limit $|k_x|\ell_T\ll 1$ we have $K_0(|k_x(y-y')|)\approx \ln(2/|k_x(y-y')|)-\gamma$,
where $\gamma$ is the Euler constant. The effect of the gate becomes 
essential if $d$ is not too large, i.e., for $2|k_{x}| d \alt 1$. 
For the gated sample and $4d^{2} \gg \ell_{T,0}^{2}$, in the long-wavelength 
limit $2|k_{x}| d \ll 1$,  we have $R_{g} \approx \ln(2d/|y-y^{\prime}|)$.

From Eq. (\ref{eq29}) it follows that $\rho^{ru}(\omega, k_x,y)$ and $\rho^{rd}(\omega, k_x,y)$ can be well approximated by
\begin{eqnarray}
&&\hspace*{-0.59cm}\rho^{ru}(\omega, k_x,y)=  
\Big[4\ell_T\cosh^2({y-y_r^u\over 2\ell_T})\Big]^{-1}\,\rho^{ru}(\omega, k_x) , \\*
&&\hspace*{-0.59cm}\rho^{rd}(\omega, k_x,y)= 
(1/\sqrt{\pi}\ell_0) e^{-(y-y_r^d)^2/  
\ell_0^2}
\,\rho^{rd}(\omega, k_x).
\label{eq3031}
\end{eqnarray}

If we assume $y_r^d - y_r^u\gg \ell_T$, we can neglect any overlap between $\rho^{ru}(\omega, k_x,y)$ and $\rho^{rd}(\omega, k_x,y)$ 
in Eq. (\ref{eq29}). Then, by  integration of Eq. (\ref{eq29}) over $y$ within separate regions around $y_{r}^{u}$ and $y_{r}^{d}$,
we obtain two coupled equations for $\rho^{ru}(\omega, k_x)$ and $\rho^{rd}(\omega, k_x)$. They read
\begin{eqnarray}
\nonumber
&&\Big[(\omega -k_xv_g^u) -2c_{h}k_x 
a_{p}(k_{x};d)\Big]
\rho^{ru}(\omega, k_x)\\* 
&&\hspace*{-0.11cm}-2c_{h}k_x  
R_{g}(|y_r^d-y_r^u|,k_{x};d) 
\rho^{rd}(\omega, k_x)=0,
\label{eq32}
\end{eqnarray}
with $v_{g}^{u}=(\ell_{0}/\sqrt{2 \pi}\Delta y) v_{F} \approx 4 \times 10^{6}$cm/s, and
\begin{eqnarray}
\nonumber
&&\Big[(\omega -k_xv_g^d)+c_{h}k_x  a_{m}(k_{x};d) \Big]
\rho^{rd}(\omega, k_x)\\*
&&\hspace*{-0.14cm}+\,c_{h}k_x  \,
R_{g}(|y_r^d-y_r^u|,k_{x};d) \rho^{ru}(\omega, k_x)=0 ,
\label{eq33}
\end{eqnarray}
with $|v_{g}^{d}| \gg v_{g}^{u}$ and $v_{g}^{d}<0$. Indeed, we 
estimate a typical $|v_{g}^{d}| \sim 3 \times 10^{7}$ cm/s  
using numerical results from, e.g., Ref.
\onlinecite{gus}. Notice that from Refs. \onlinecite{bask,per} and Sec. II we 
obtain $|v_{g}^{d}|<v_{F} \approx 10^{8}$cm/s; that is, the group velocity 
of any edge state must be smaller than $v_{F}$. 
The matrix elements $a_{p}(k_{x};d)$  and $a_{m}(k_{x};d)$ are given by
\begin{equation}
\hspace*{-0.3cm}a_{p}(k_{x};d)=\frac{1}{16} \int_{-\infty}^{\infty} \int_{-\infty}^{\infty}
\frac{dx dt \; R_{g}(\ell_{T}|x-t|,k_{x};d)}{\cosh^{2}(x/2)\cosh^{2}(t/2)}  ,
\label{eq34a}
\end{equation}
\begin{equation}
a_{m}(k_{x};d)=\frac{1}{\pi} \int_{-\infty}^{\infty} \int_{-\infty}^{\infty} \frac{dx dt}{e^{x^2+t^2}} 
R_{g}(\ell_{0}|x-t|,k_{x};d) .
\label{eq34b}
\end{equation}
\begin{figure}[ht]
\vspace*{-0.5cm}
\includegraphics [height=11cm, width=8cm]{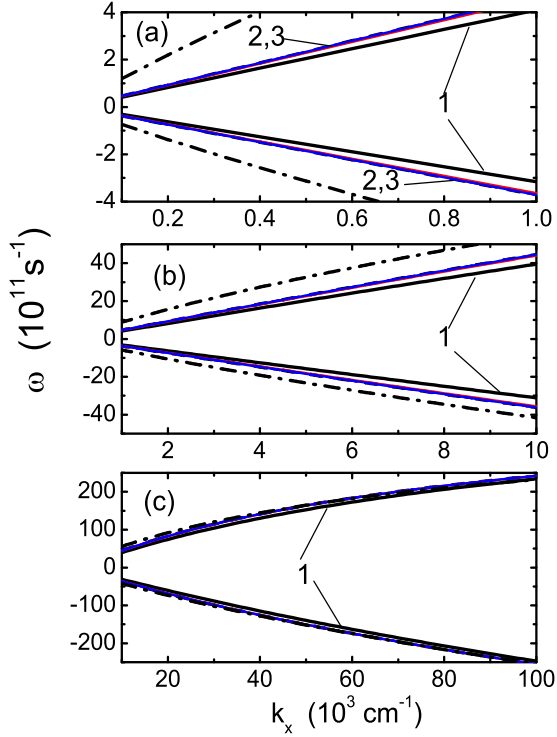} 
\vspace*{-0.7cm}
\caption{(Color online) The dispersion relations $\omega_{\pm}^{(i)}(k_x,d=300$ nm) (solid curves),
$\omega_{\pm,0}^{(i)}(k_x,d=300$ nm) (dashed curves),
and $\omega_{\pm,0}^{(i)}(k_x,d \to \infty)$ (dot-dashed curves) of two counter 
propagating fundamental EMPs for  case (i) at $\nu=2$. Panels (a), (b), and (c)
correspond to  three characteristic $k_x$ regions: 
$10^{3}$\,cm$^{-1}\geq k_x \geq 10^{2}$\, cm$^{-1}$ in (a),
$10^{4}$\, cm$^{-1} \geq k_x \geq 10^{3}$\, cm$^{-1}$ in (b), and
$10^{5}$\, cm$^{-1}\geq k_x \geq 10^{4}$\, cm$^{-1}$ in (c).
The solid curves marked by 1, 2, and 3 correspond, respectively, to 
inter-edge distances, in the $n=0$ LL, $y_{r}^{d}-y_{r}^{u}=\Delta y$, $4 \Delta y$, and 
$16 \Delta y$.
The other parameters are $B=9$T, $T=77$K, $\ell_{T}/\ell_{0}=2$,  $\Delta y=10 \ell_{0}$,
$v_{g}^{u}=4 \times 10^{6}$ cm/s, $v_{g}^{d}=-3 \times 10^{7}$ cm/s, $\epsilon=2$, and      $\ell_{0}\approx 8.5$ nm. }
\end{figure}

For $|k_x(y_r^d-y_r^u)| \gg 1$ it's a good approximation to neglect the terms $\propto R_g(...)$ in Eqs. (\ref{eq32}) and  
(\ref{eq33}). Then  Eqs. (\ref{eq32}) and  (\ref{eq33}) are decoupled.  The resulting  
dispersion relations for the two fundamental EMP 
modes are
\begin{equation}
\omega_{+,0}^{(i)}(k_x,d)=k_x v_g^u+ 2c_{h}k_x  a_{p}(k_{x};d) ,
\label{eq34}
\end{equation}
for the mode localized at $y_{r}^{u}$, 
that has {\it positive} phase and group velocities, 
and
\begin{equation}
\omega_{-,0}^{(i)}(k_x,d)=-k_x|v_g^d|-c_{h}k_x  \, a_{m}(k_{x};d) ,
\label{eq35}
\end{equation}
for the mode localized at $y_{r}^{d}$ that has {\it negative} phase and group velocities.
Notice that  in the long-wavelength limit and for large $d$ the effect of 
the gate, $\propto \exp(-2|k_{x}| d) \ll 1$,  can be neglected in Eqs. (\ref{eq34}) and (\ref{eq35}). The result is  
$a_{p}(k_{x};d) \to \left[\ln(1/|k_x|\ell_T)-0.145\right]$ and  
$a_{m}(k_{x};d) \to \left[\ln(1/|k_x|\ell_0)+3/4\right]$. 

\begin{figure}[ht]
\vspace*{-0.5cm}
\includegraphics [height=10cm, width=8cm]{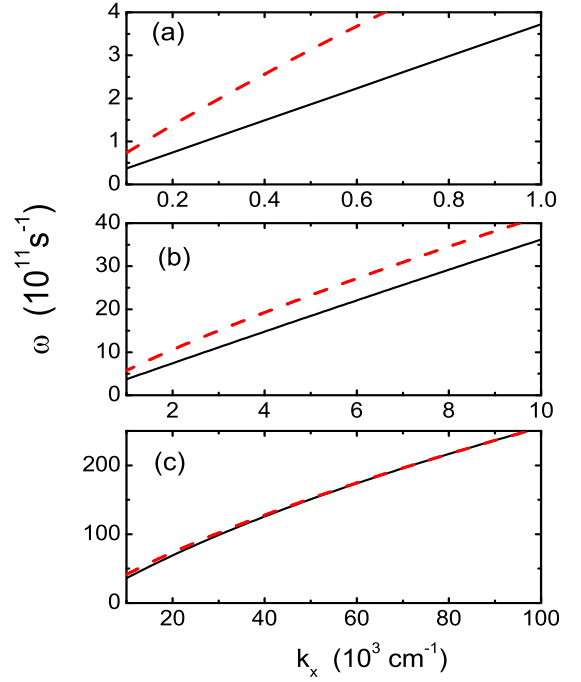} 
\vspace*{-0.9cm}
\caption{(Color online)    The dispersion relations of the unique, left fundamental 
EMP for case (ii) and  two values of $d$,
$\omega^{(ii)}(k_x,d=300$ nm) (solid curves), and $\omega^{(ii)}(k_x,d=\infty)$ (dashed curves).  
Panels (a), (b), and (c) correspond, respectively, 
to panels (a), (b), and (c) of Fig. 2. 
Here  $v_{g}^{u1}=3 \times 10^{7}$ cm/s and the other parameters are the same as those in Fig. 2. } 
\end{figure}

For  case (ii),  in the low-frequency limit $\omega\ll v_F/\ell_0$, the result is 
\begin{equation}
\rho^{r,u1}(\omega, k_x,y)= {1\over
\sqrt{\pi}\ell_0} e^{-(y-y_r^{u,1})^2/\ell_0^2}\,\rho^{r,u1}(\omega, k_x).
\label{eq36}
\end{equation}
Using Eq.  (\ref{eq25}) and other relevant expressions gives, for $\rho^{r,u1}(\omega, k_x) \neq 0$, 
the dispersion relation for only one fundamental EMP mode, localized mainly at $y_{r}^{u,1}$,
\begin{equation}
\omega^{(ii)}(k_x,d)=k_xv_g^{u1}+c_{h}k_x  \, a_{m}(k_{x};d) ,
\label{eq37}
\end{equation}
with {\it positive} phase and group velocities.  Here $v_{g}^{u1}>0$ and, 
similar to  $|v_{g}^{d}|$, we estimate  $v_{g}^{u1} \alt 3 \times 10^{7}$ cm/s. 

If we take into account the coupling in Eqs. (\ref{eq32})-(\ref{eq33}), due 
to $R_{g}(|y_r^d-y_r^u|,k_{x};d) \neq 0$, then a nontrivial solution of this 
system requires its determinant to vanish. This leads to two renormalized EMP modes,  $\omega^{(i)}_{+}$ 
and $\omega^{(i)}_{-}$, 
\begin{eqnarray}
\nonumber
&&\omega_{\pm}^{(i)}(k_x,d)=\frac{1}{2}\left[\omega_{+,0}^{(i)}(k_x,d)+\omega_{-,0}^{(i)}(k_x,d)\right] \\*
&&\pm \frac{1}{2} 
\Big[\big[\omega_{+,0}^{(i)}(k_x,d)-\omega_{-,0}^{(i)}(k_x,d)\big]^{2}  \nonumber \\*
&& -8 c_{h}^2\, k_x^2\,  R_{g}^{2}(y_{r}^{d}-y_{r}^{u},k_{x};d)\Big]^{1/2},
\label{eq38}
\end{eqnarray}
where $\omega_{\pm,0}^{(i)}(k_x,d)$ are given by Eqs. (\ref{eq34})-(\ref{eq35}). 
If we neglect the Coulomb coupling $R_{g}(...)$ between the
charge excitations at $y_r^d$ and $y_r^u$,  Eq. (\ref{eq38}) leads to the limits 
$\omega_{+}^{(i)}(k_x,d) \to \omega_{+,0}^{(i)}(k_x,d)$
and $\omega_{-}^{(i)}(k_x,d) \to \omega_{-,0}^{(i)}(k_x,d)$.

From Eqs. (\ref{eq38}) and (\ref{eq32})-(\ref{eq33}) it follows that
\begin{eqnarray}
\nonumber
&&\rho^{ru}(\omega_{+}^{(i)}(k_x,d),k_x)/\rho^{rd}(\omega_{+}^{(i)}(k_x,d),k_x) \\*
&&=2 \rho^{rd}(\omega_{-}^{(i)}(k_x,d),k_x)/\rho^{ru}(\omega_{-}^{(i)}(k_x,d),k_x) ,
\label{eq39}
\end{eqnarray}
for any $d$, $y_{r}^{d}-y_{r}^{u}$, and $k_{x}$, in particular  for $d \to \infty$. 
That is, the ratio of the charge amplitudes 
$\rho^{ru}(\omega_{+}^{(i)}(k_x,d),k_x)/\rho^{rd}(\omega_{+}^{(i)}(k_x,d),k_x) \equiv \rho^{ru}_{+}/\rho^{rd}_{+}$ 
for the $\omega_{+}^{(i)}(k_x,d)$ EMP, at the edges $y_{r}^{u}$ and $y_{r}^{d}$, times  the   
ratio $\rho^{ru}_{-}/\rho^{rd}_{-}$, for the $\omega_{-}^{(i)}(k_x,d)$  EMP, is equal to $2$.

For  case (i) and $\nu=2$, in Fig. 2 we plot the dispersion relations $\omega_{\pm}^{(i)}(k_x,d=300$ nm) (solid curves, Eq. (\ref{eq38})),
$\omega_{\pm,0}^{(i)}(k_x,d=300$ nm) (dashed curves, Eqs. (\ref{eq34})-(\ref{eq35})), and 
$\omega_{\pm,0}^{(i)}(k_x,d \to \infty)$ (dot-dashed curves) 
for $v_{g}^{u}=4 \times 10^{6}$ cm/s, $v_{g}^{d}=-3 \times 10^{7}$ cm/s, and $\epsilon=2$, 
in three characteristic $k_x$ regions: $10^{3}$\,cm$^{-1}\geq k_x \geq 10^{2}$\, cm$^{-1}$ in (a),
$10^{4}$\,cm$^{-1} \geq k_x \geq 10^{3}$\, cm$^{-1}$ in (b), and $10^{5}$\,cm$^{-1}\geq k_x \geq 10^{4}\,$ cm$^{-1}$ in (c). Here we 
assume that on one side of the graphene sheet there is SiO$_{2}$ substrate, 
with  dielectric constant $\approx 3$, 
and on the other side there is  air or vacuum: then for $\epsilon$ we must use, in all  formulas, 
an effective dielectric constant $\approx 2$. The other parameters used are $B=9$T, $T=77$K,  
$\Delta y=10 \ell_{0}$,  which gives $\ell_{T}/\ell_{0}=2$, and
$\ell_{0} \approx 8.5$ nm. The solid curves marked by 1, 2, and 3 
correspond to the inter-edge distance of the $n=0$ LL $y_{r}^{d}-y_{r}^{u}=\Delta y$, 
$4 \Delta y$, and $16 \Delta y$, respectively.
For any of these curves we assume $(y_{r}^{d}-y_{r}^{u})/y_{r}^{d} \ll 1$;
 the dashed and  dot-dashed curves are independent of the
inter-edge distance.  This allows us to neglect the coupling of the 
fundamental EMPs,  localized in some regions of $y>0$, with
any EMPs on the left part of channel. In particular, for $y_{r}^{d}-y_{r}^{u}=16 \Delta y$ the  
channel width is much larger than that used in Fig. 1. The solid curves 1, 2, and 3 
are very close in (a) to pertinent dashed curve  
and are even closer in (b) and (c). Resonances due to these two counter propagating
EMPs (localized in a region of extent $\leq 1\mu$m, from the right edge  at $y \approx y_{r}^{d}$)
are possible in (a),(b), and (c) for $L_{x} \sim  10^{-2}$cm,  $\sim 10^{-3}$ cm, and 
$\sim 10^{-4}$ cm, respectively.  In Fig. 2(a) the solid and dashed curves 
show a strong effect of the gate, compare with  the dot-dashed curves,
and their behavior is very close to a linear one.  In Fig. 2(c) both the effect 
of the gate and that of $y_{r}^{d}-y_{r}^{u}$ become 
essentially smaller for the curves 1, 2, and 3. 
\begin{figure}[ht]
\vspace*{-1cm}
\includegraphics [height=10cm, width=8cm]{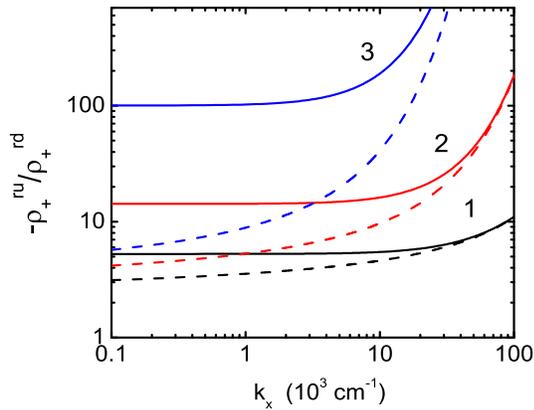} 
\vspace*{-3.9cm}
\caption{(Color online) The ratio $-\rho_{+}^{ru}/\rho_{+}^{rd} = -2\rho_{-}^{rd}/\rho_{-}^{ru}$ versus $k_x$ for  case (i), $\nu=2$, 
and other conditions as in Fig. 2.  
The solid (dashed) curves correspond to $d=300$ nm ($d \to \infty$).  The curves $1$, $2$, and $3$ 
correspond to $y_r^d-y_r^u=\Delta y$, $4\Delta y$, and $16\Delta y$, respectively, and $\Delta y=10 \ell_{0}$.}
\end{figure}

For  case (ii) and $\nu=2$  in Fig. 3 we plot the dispersion relations of 
the unique left fundamental EMP $\omega^{(ii)}(k_x,d)$, Eq. (\ref{eq37}), for two 
characteristic values of $d$,  $d=300$ nm (solid curves) and $d=\infty$ (dashed curves),
and the same three characteristic $k_x$ regions of Fig. 2.

\begin{figure}[ht]
\vspace*{-1cm}
\includegraphics [height=10cm, width=8cm]{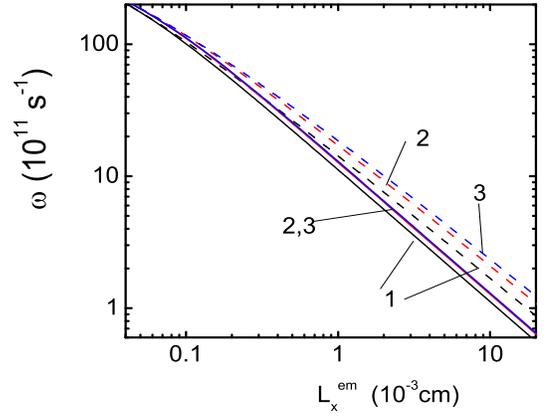}
\vspace*{-3.5cm}
\caption{(Color online) The main resonance frequency $\omega$ as function of $L_{x}^{em}$, 
calculated from Eq. (\ref{eq40}) with $N=1$, for  case (i), $\nu=2$, and other conditions as in Fig. 2.  
The resonance is due to two counter propagating fundamental EMPs, Eq. (\ref{eq38}), localized  
between the edge states, at $y_{r}^{u}$ and $y_{r}^{d}$, of the $n=0$ LL. 
The solid (dashed) curves correspond to $d=300$ nm ($d \to \infty$).  
The curves $1$, $2$, and $3$ correspond to $y_r^d-y_r^u=\Delta y$, $4\Delta y$, and $16\Delta y$, 
respectively, $ \Delta y=10 \ell_{0}$, and $\ell_{0}=8.5$ nm.}
\end{figure}

In Fig. 4, for  case (i) and other conditions as in Fig. 2, 
we plot the ratio $-\rho_{+}^{ru}/\rho_{+}^{rd}$ versus $k_x$. The curves $1$, $2$,  and $3$ 
are obtained from Eqs. (\ref{eq32}), (\ref{eq38})-(\ref{eq39}).
The solid curves correspond to $d=300$ nm and the dashed ones to $d \to \infty$.  
According to Eq. (\ref{eq39}) we have $\rho_{+}^{ru}/\rho_{+}^{rd}= 2\rho_{-}^{rd}/\rho_{-}^{ru}$. 
Figure 4 shows that the fundamental EMPs  with positive 
($\omega_{+}^{(i)}(k_x,d)/k_{x}>0$), and negative phase velocity ($\omega_{-}^{(i)}(k_x,d)/k_{x}<0$),  
renormalized by the 
inter-edge Coulomb interaction, have their charge density amplitudes, 
at different edges (i.e., at $y_{r}^{u}$ and $y_{r}^{d}$),
in opposite phase. Moreover, for the former EMP the charge 
excitation is mainly localized at the edge 
$y_{r}^{u}$ (i.e., the position of edge states due to the smooth confining 
potential) as $-\rho_{+}^{ru}/\rho_{+}^{rd}  >1$, whereas   for the latter EMP it is mainly localized at the edge 
$y_{r}^{d}$ as $-\rho_{-}^{ru}/\rho_{-}^{rd}  <1$. 

In Fig. 5, for  case (i) and other conditions as in Fig. 2, we plot the main resonance, at $N=1$, 
obtained for the $\omega_{\pm}^{(i)}(k_x,d)$ EMPs from 
\begin{equation}
\left[k_{x}^{+}(\omega)-k_{x}^{-}(\omega)\right]L_{x}^{em}=2\pi N ,
\label{eq40}
\end{equation}
where $L_{x}^{em}$ is the length of a segment  of  the right edge along which the EMPs  
propagate freely, see Eq. (\ref{eq38}).  Due to the counter propagation of these 
two EMPs, the relation $L_{x}^{em} \leq L_{x}$ is possible. In particular, 
we will have $L_{x}^{em} \ll L_{x}$ if a strong coupling between 
the EMPs, Eq. (\ref{eq38}), is introduced in the relevant high-frequency range 
within two spatial regions separated by $L_{x}^{em} \ll L_{x}$,  
along the right graphene edge(s). Actually, as we see, e.g., in Fig. 1(a) the right edge in   case (i)   
consists of  two edges, at $y_{r}^{u}$ and $y_{r}^{d}$, 
with pertinent edge states (due to two intersections of the $n=0$ LL
with the Fermi level). Despite that $y_{r}^{d}$ is very close to
the right armchair termination of the channel, it follows that $L_{y}/2>y_{r}^{d}$.
In Eq. (\ref{eq40})  the EMP's wave vector  $k_{x}^{\pm}(\omega)$
is obtained from Eq. (\ref{eq38}) abbreviated as $\omega=\omega_{\pm}^{(i)}(k_{x}^{\pm},d)$. 

In addition, for case (i) a strong Bragg coupling is possible due to a weak superlattice along 
the edge, with period $L_{x}^{em}$, if $L_{x}/L_{x}^{em} \gg 1$. In particular, for 
frequencies in the THz range, see Fig. 5, 
$L_{x}^{em} \alt 1 \mu$m and 
$L_{x}/L_{x}^{em} \agt 10^{2}$ correspond to rather typical lengths 
 $L_{x} \agt 10^{2} \; \mu$m in experiments.

\section{ Concluding Remarks }

We studied EMPs near an armchair edge of a wide graphene channel,
at $y=L_{y}/2$,  with a smooth lateral potential, in the $\nu=2$ regime of  QHE and 
when the Fermi level $E_{F}$ is in a gap.  We showed that the position of $E_{F}$ 
can strongly affect the chirality,  spectrum,  spatial structure, and the 
 number of  the fundamental EMPs. When $E_{F}^{(1)}$ intersects four degenerate 
states of the $n=0$ LL at $y_{r}^{u}>0$
and two degenerate states of this level at $y_{r}^{d} \gg y_{r}^{d}-y_{r}^{u} \gg \ell_{0}$ 
(case (i)), two fundamental EMPs, with opposite chirality, 
counter propagate along the right edge of the channel.
For the same wave vector the absolute values of their phase velocities are 
different and they have spatial structure along the $y$ axis, 
with an essential overlap in the region
between the edge states, at $y_{r}^{u}$ and  $y_{r}^{d}$, and their vicinity.
That is, the right edge consists of two edges at $y_{r}^{u}$ and $y_{r}^{d}$,
with pertinent edge states, due to the $n=0$ LL.
When the Fermi level is 
sufficiently high, $E_{F}^{(2)}$, and intersects only two degenerate states  
of the $n=0$ LL at $y_{r}^{u,1} \approx y_{r}^{d}$ (case (ii)), 
only one fundamental EMP exists,  of the "usual'' chirality
for edges of $n$ type conventional 2DES. 

In case (i) we found that a resonance of two fundamental EMPs, of opposite 
chirality, on the (right) edge of a graphene channel, is possible in a wide region of frequencies.
The main resonance described by Eq. (\ref{eq40}) is allowed in 
segment lengths $L_{x}^{em} \leq L_{x}$  along the edge.
The $N=1$ resonance means that the sum of the total increases of the wave 
phases  of the $\omega_{+}^{(i)}(k_x,d)$ and $\omega_{-}^{(i)}(k_x,d)$ EMPs, during their propagation between the ends of  
$L_{x}^{em}$ along  the positive and   negative 
$x$ axis, respectively, is equal to $2\pi$. This partly resembles the condition for the main 
resonance of a usual EMP, see, e.g., Ref. \cite{fetter,volkov91}, where the EMP  propagating along the perimeter 
$P$ of a conventional 2DES, typically  $P \agt 10^{-1}$cm) \cite{fetter,volkov91,bal}, acquires a phase $2\pi$. 
Moreover, for experimentally  realistic values of $L_{x}$ we obtained $L_{x}^{em} \ll L_{x}$. 
Indeed, for  frequencies in the THz range and $L_{x}^{em} \alt 1 \mu$m the  experimentally 
realistic values $L_{x} \sim 10^{2} \mu$m entail 
$L_{x}/L_{x}^{em} \geq 10^{2}$. Then we can speculate that a strong Bragg coupling is possible  
due to a weak periodic superlattice along the edge with   period $L_{x}^{em}$. Notice that 
a weak  superlattice potential along the edge, with  period $\agt 10^{2}$ nm, has 
negligible effect on a fundamental EMP in the QHE regime in  conventional 2DESs \cite{bal2000b}. 
In addition, as two renormalized fundamental EMP modes  $\omega^{(i)}_{\pm}$, Eq. (\ref{eq38}), 
are counter propagating and their spatial structures have essential overlap along $y$,
in a narrow region between $y_{r}^{u}$ and $y_{r}^{d}$, that can easily be modified due 
to the strong dependence of  $y_{r}^{u}$ on a smooth lateral electrostatic potential,
time-resolved experiments, such as those of Refs. \onlinecite{ashoori92,ernst96}, can be 
used to observe appearance of counter propagating EMP along the 
armchair edge in the $\nu=2$ QHE regime. As far as, shown in Fig. 1(a), case (i) is realized, 
that can be realized for a wide range of parameters.
The entire EMP picture and properties are different from those of EMPs 
in  conventional 2DESs due to the difference in the spectrum of the edges
and the corresponding wave functions. 
In fact, as was mentioned earlier in Fig. 1(b) and above Eq. (17),   
in the absence of a smooth electrostatic potential the spectrum  agrees with 
the usual, hard-wall potential of Refs. \onlinecite{brey,aba,gus0,gus}. 
Correspondingly, we don't have two counterpropagating fundamental EMPs but 
only one fundamental EMP with properties similar to those of 
the fundamental EMP in conventional 2DES.

Next we list and discuss the approximations used. In  studying the EMPs in the $\nu=2$ QHE 
regime we neglected  dissipation. This approximation is well justified as the EMP damping can be
related  only with inelastic scattering processes within narrow 
temperature belts, of width $k_BT$, of each edge state, cf.  \cite{bal,bal2000}, 
that are much weaker than  scattering processes
due to a static disorder, especially in the QHE regime which implies relatively low $T$. 
The latter  makes a dominant contribution to the transport scattering time in a 2DES of 
graphene \cite{novo,cast,vasko09} for $B=0$. 
Further, the damping of the EMPs will influence some properties of a Bragg coupling and 
the quality of the EMP resonances, cf. Eq. (\ref{eq40}). Notice that for decreasing 
temperature $T$ any EMP damping will quickly weaken. However,  for sufficiently small 
$T$ the condition $\ell_{T}/\ell_{0} \gg 1$ can be violated. Nevertheless, at quite 
low $T$ and for sufficiently smooth bare confining potential,
the group velocity can essentially decrease with decreasing temperature \cite{bal2000} due to many-body effects.
In addition, even for $\ell_{T}/\ell_{0} \ll 1$
it appears that the present results will be only weakly and quantitatively  
modified since the   contributions to a fundamental EMP coming from a region of the LL edge, at $y_{r}^{u}$, 
will bring about only small changes \cite{bal,bal2000,bal99}. Obviously, for a more accurate account of 
the EMPs studied here  dissipation must be included in the treatment. We emphasize that our study of the 
fundamental EMPs for the armchair termination of a graphene channel
cannot be directly extended to  zigzag termination as some important properties of the wave functions and the
energy levels are different than those of the armchair termination,  cf. \cite{cast,brey,aba,gus0,gus}. 
We relegate the study of EMPs along zigzag edges to a future work.

We used a simple analytical model of a smooth, lateral confining potential Eq. (\ref{eq17}), but our main 
results are quite robust to modifications of its form and parameters since 
cases (i) and (ii) can be realized in a graphene channel in the $\nu=2$ QHE regime. 
Further, near  the edge states at $y_{r}^{d}$ and $y_{r}^{u,1}$ we used a simple analytic model to 
approximate a static density profile, cf.  Eq. (\ref{eq23}). 
In doing so we neglected possible modifications of the static density profile due to local
charging \cite{thouless93,macdonald83,thouless85} $\propto d^{2}V(y)/dy^{2}$. 
Notice that these modifications are weak for a smooth potential  and their neglect 
should have a minor effect in the fundamental EMPs we studied as their main properties 
are very robust against  details of  a static density profile\cite{bal,bal2000,bal99} and, 
in particular,   nonlocal effects  \cite{bal99}.

\begin{acknowledgments}
This work was supported by the Brazilian Council for Research (CNPq) APV Grant No. 452849/2009-8  and the Canadian 
NSERC Grant No. OGP0121756, O. G. B. also acknowledges support by Brazilian 
FAPEAM  (Funda\c{c}\~{a}o de Amparo \`{a} Pesquisa do Estado do
Amazonas) Grant.
\end{acknowledgments}
\appendix
%


%

\begin{thebibliography}{10}
%
%
\bibitem{novo} K. S.~Novoselov, A. K.~Geim, S. V.~Morozov, D.~Jiang, Y.~Zhang, S. V.~Dubonos, I. V.~Grigorieva, and A. A.~Firsov, Science {\bf 306}, 666 (2004);
K. S.~Novoselov,   {\it Proc. Natl. Acad. Sci. USA} {\bf 102}, 10451 (2005);
A. K.~Geim and K. S.~Novoselov, Nature Materials, {\bf 6}, 183 (2007). 


\bibitem{cast} A. H.~Castro Neto, F.~Guinea, N. M. R.~Peres, K. S.~Novoselov, and A. K.~Geim, Rev.~Mod.~Phys.~{\bf 81}, 109 (2009).

\bibitem{wallace} P. R. Wallace,   Phys. Rev. {\bf 71}, 622 (1947).

\bibitem{klein} O. Klein, Z. Phys. {\bf 53}, 157 (1929).

\bibitem{kat} M. I. Katsnelson, K. S. Novoselov, A. K. Geim,
Nature Phys. {\bf 2}, 620 (2006); J. Milton Pereira Jr.,  P. Vasilopoulos, and F.
M. Peeters, Appl. Phys. Lett. {\bf 90}, 132122, (2007).

\bibitem{brey} L. Brey and H. A. Fertig,   Phys. Rev. B {\bf 73}, 195408 (2006);
N. M. R. Peres, F. Guinea, A. H. Castro Neto, {\it ibid} {\bf 73}, 125411 (2006).

\bibitem{aba} D. A. Abanin, P. A. Lee, and L. S. Levitov, Phys. Rev. Lett. {\bf 96}, 176803 (2006);
Solid State Commun., {\bf 143}, 77 (2007).

\bibitem{gus0} V. P. Gusynin, V. A. Miransky,  S. G. Sharapov, and I. A. Shovkovy,  Phys. Rev. B {\bf 77}, 205409 (2008).

\bibitem{zit1} N. M. R. Peres,  A. H. Castro Neto, F. Guinea, Phys. Rev. B {\bf 73}, 241403 (2006).


\bibitem{zit2} H.-Y. Chen, V. Apalkov, and T. Chakraborty, Phys. Rev. Lett.  {\bf 98}, 186803 (2007);
 A. V. Shytov, M. S. Rudner, and L. S. Levitov, {\it ibid} {\bf 101}, 156804 (2008). 


\bibitem{pom} L. A. Ponomarenko {\it et al.}, Science {\bf 320}, 356 (2008); 
C. Stampfer, E. Schurtenberger, F. Molitor, J. Guttinger, T. Ihn, and K. Ensslin, Nano Lett.  {\bf 8}, 2378 (2009).


 \bibitem{loss} B. Trauzettel  Denis V. Bulaev, D. Loss, and G. Burkard, Nature. Phys. {\bf 3}, 192 (2007);
K. Nakada, M. Fujita, G. Dresselhaus, and M. S. Dresselhaus, Phys. Rev. B {\bf 54}, 17954 (1996).

\bibitem{gus} V. P. Gusynin, V. A. Miransky,  S. G. Sharapov, I. A. Shovkovy, and C. M. Wyenberg,  Phys. Rev. B {\bf 79}, 115431 (2009).
 
\bibitem{milt} J.M. Pereira, F. M. Peeters, and P.~Vasilopoulos, Phys. Rev. B {\bf 75}, 125433 (2007). 

\bibitem{plasmon} S. Das Sarma and E. H. Hwang, Phys. Rev. Lett. {\bf 102}, 206412 (2009);
M. Polini, R. Asgari, G. Borghi, Y. Barlas, T. Pereg-Barnea, and A. H. MacDonald, 
Phys. Rev. B {\bf 77}, 081411(R) (2008);
Yu Liu, R. F. Willis, K. V. Emtsev, and Th. Seyller, 
{\it ibid} {\bf 78}, 201403(R) (2008).


\bibitem{xue} B. Wunsch, T. Stauber, F. Sols, and F. Guinea, New J. Phys, {\bf 8}, 318 (2006);
X. F. Wang and T. Chakraborty, Phys. Rev. B {\bf 75}, 033408 (2007);.
T. Langer, J. Baringhaus, H. Pfnur, H. W. Schumacher,
and C. Tegenkamp, New J. Phys, {\bf 12}, 033017 (2010).

\bibitem{magnplasmon} O. L. Birman, G. Gumbs, and Y. E. Lozovik, Phys. Rev. B {\bf 78}, 085401 (2008); 
R. Roldan, J.-N. Fuchs, and M. O. Goerbig, {\it ibid}  {\bf 80} 085408 (2009); Yu. A. Bychkov and G. Martinez,
{\it ibid}  {\bf 77} 125417 (2008); A. Iyengar, J. Wang, H. A. Fertig, and L. Brey, {\it ibid}  {\bf 75} 125430 (2007).


\bibitem{volkov91} V.A. Volkov and S.A. Mikhailov, ``Electrodynamics of Two-Dimensional Electron Systems in 
High Magnetic Fields,'' in Landau Level Spectroscopy, Modern Problems in Condensed Matter Sciences,
Ed. by G. Landwehr and E. I. Rashba (North-Holland, Amsterdam,
1991), vol. 27.2, ch.15, p. 855-907; V.A. Volkov and S.A. Mikhailov, Zh. Eksp. Teor. Fiz. {\bf 94}, 217 (1988)
[Sov. Phys. JETP {\bf 67}, 1639 (1988)].

\bibitem{kushwaha2001} M.S. Kushwaha, Surface Science Reports {\bf 41}, p. 1-416 (2001).

\bibitem{fetter} A. L. Fetter, Phys. Rev. B {\bf 32}, 7676 (1985); V. A. Volkov and S. A. Mikhailov,
Pis'ma Zh. Eksp. Teor. Fiz. {\bf 42}, 450 (1985) [JETP Lett.{\bf 42}, 556 (1985)].

\bibitem{aleiner94} I. L. Aleiner and L. I. Glazman, Phys. Rev. Lett. {\bf 72}, 2935 (1994).


\bibitem{wen91} B. I. Halperin, Phys. Rev. B {\bf 25}, 2185 (1982); 
X. G. Wen, {\it ibid} {\bf 43}, 11025 (1991); 
M. Stone, Ann. Phys. (N.Y.) {\bf 207}, 38 (1991).

\bibitem{stone92} M. Stone, H. W. Wyld, and R. L. Shult, Phys. Rev. B {\bf 45}, 14156 (1992);
U. Zulicke and A. H. MacDonald, {\it ibid} {\bf 54}, 16813 (1996);
S. Giovanazzi,L. Pitaevskii, and S. Stringari, Phys. Rev. Lett. {\bf 72}, 3230 (1994).

\bibitem{bal} O. G. Balev and P.~Vasilopoulos, Phys. Rev. Lett. {\bf 81}, 1481 (1998); O.G. Balev, P. Vasilopoulos, and
Nelson Studart, J. Phys.: Condens. Matter {\bf 11}, 5143 (1999); 
O. G. Balev and P.~Vasilopoulos, Phys. Rev. B {\bf 56}, 13252 (1997). 

\bibitem{bal2000} O. G. Balev and Nelson Studart, Phys. Rev. B {\bf 61}, 2703 (2000);
Sanderson Silva and O. G. Balev, J. Appl. Phys. {\bf 107}, 104310 (2010); I.O. Baleva,
N. Studart, and O.G. Balev, Phys. Rev. B {\bf 65}, 073305 (2002). 


\bibitem{bal99} O. G. Balev and P.~Vasilopoulos, Phys. Rev. B {\bf 59}, 2807 (1999). 

\bibitem{bal2000b} O. G. Balev,  Nelson Studart, and P. Vasilopoulos, Phys. Rev. B {\bf 62}, 15834 (2000). 



\bibitem{ashoori92} R. C. Ashoori, H. L. Stormer, L. N. Pfeiffer, K. W. Baldwin, and K. West, 
Phys. Rev. B {\bf 45}, 3894 (1992).

\bibitem{ernst96} G. Ernst, R. J. Haug, J. Kuhl, K. von Klitzing, and K. Eberl, Phys. Rev. Lett. {\bf 77}, 4245 (1996).

\bibitem{kukushkin09} M. N. Khannanov, A. A. Fortunatov, and I. V. Kukushkin, 
Pis'ma Zh. Eksp. Teor. Fiz. {\bf 90}, 740 (2009) [JETP Lett.{\bf 90}, 667 (2009)].
   

\bibitem{bask} V. Lukose, R. Shankar, and G. Baskaran, Phys. Rev. Lett. {\bf 98}, 116802 (2007).  


\bibitem{per}N. M. R.~Peres and E. V. Castro, J. Phys. Condens. Matter {\bf 19}, 406231 (2007). 


\bibitem{bee} C. W. J. Beenakker and H. van Houten, in {\it Quantum Transport in
Semiconductor Nanostructures}, Solid State Physis Vol. 44 
edited by H. Ehrenreich and D. Turnbull (Academic, San Diego, 1991). 

\bibitem{thouless93} D. J. Thouless, Phys. Rev. Lett. {\bf 71}, 1879 (1993).

\bibitem{bal96} O. G. Balev and P.~Vasilopoulos, Phys. Rev. B {\bf 54}, 4863 (1996). 


\bibitem{zhe} Y. Zheng and T. Ando, Phys. Rev. B {\bf 65}, 245420 (2002); V. P. Gusynin and S. G. Sharapov,
Phys. Rev. Lett. {\bf 95}, 146801 (2005). 

\bibitem{vasko09} O. G. Balev, F. T. Vasko, and V. Ryzhii, Phys. Rev. B {\bf 79}, 165432 (2009).

\bibitem{macdonald83} A. H. MacDonald, T. M. Rice, and W. F. Brinkman, Phys. Rev. B {\bf 28}, 3648 (1983).

\bibitem{thouless85} D. J. Thouless, J. Phys. C {\bf 18}, 6211 (1985).

\end{thebibliography}
\end{document}